\documentstyle[12pt]{article}
\def\x{{\mbox{\boldmath$x$}}}
\def\u{{\mbox{\boldmath$u$}}}

\def\k{{\mbox{\boldmath$k$}}}

\def\nab{{\mbox{\boldmath$\nabla$}}}

\newcommand{\aleq}{\mbox{\ \raisebox{-.9ex}{$\stackrel{\textstyle <}{\sim}$}\
}}

\def\rel{{Re_\lambda}}
\def\eps{{\epsilon}}
\def\om{{\omega}}
\def\epsu{{\epsilon_u}}
\def\epshel{{\epsilon_{Hel}}}
\def\epst{{\epsilon_\Theta}}
\def\ptos{{P_\Theta (\omega) \propto\omega^{-1.3\pm 0.1}}}
\def\pto{{P_\Theta (\omega)}}

\def\begineq{\begin{equation}}
\def\endeq{\end{equation}}

\begin{document}
\bibliographystyle{prsty}
\title{Temperature spectra in shear flow and\\ thermal convection}
\author{Detlef Lohse $^{*}$}
\maketitle

\centerline{The James Franck Institute, The University of Chicago,}
\centerline{ 5640 South Ellis Avenue, Chicago, IL 60637, USA}

\bigskip

\date{}
\maketitle
\bigskip
\bigskip
\bigskip

We show that
the $P_u(\om) \propto \om^{-7/3}$ shear velocity power spectrum gives
rise to a $P_\Theta (\om ) \propto \om^{-4/3}$ power spectrum for a
passively advected scalar, as measured in experiment
[K. Sreenivasan, Proc. R. Soc. London A {\bf 434},  165  (1991)].
Applying our argument to high Rayleigh number Rayleigh Benard flow, we
can account for the measured scaling exponents
equally well as the
Bolgiano Obukhov theory (BO59). Yet, of the two explanations,
only the shear approach might be
able to explain why no classical scaling range is seen in between the
shear (or BO59) range and the viscous subrange of the experimental
temperature spectrum
[I. Procaccia {\it et~al.}, Phys. Rev. A {\bf 44},  8091  (1991)].


\vspace{0.5cm}\noindent


\newpage
\section{Velocity shear spectra}
The classical scaling laws in fully developed turbulence are the
Kolmogorov \cite{kol41} and the Obukhov-Corrsin \cite{obu49} scaling
relations
\begineq
E_u(k)  \sim \epsu^{2/3}  k^{-5/3}, \qquad
E_\Theta(k) \sim \epst \epsu^{-1/3} k^{-5/3}
\label{eq1}
\endeq
for the velocity spectrum $E_u (k)$ and the temperature spectrum
$E_\Theta (k)$ of a passively advected temperature field. Here,
$\epsu$ and $\epst$ are the energy and thermal dissipation rates,
respectively. Eqs.\ (\ref{eq1}) follow from dimensional analysis. They
can also be derived from the assumption that the energy flux
$T_u(k)\sim ku^3(k)$ of the conserved total energy $\int \u^2 (\x)
d\x/2$ and the corresponding flux $T_\Theta (k) \sim k u(k) \Theta^2
(k)$ of the conserved total temperature intensity $\int \Theta^2 (\x)
d\x$ do not depend on $k$, i.e., $T_u(k) \sim \epsu$ and $T_\Theta(k) \sim
\epst$. Here we have assumed that both velocity and temperature fields are
stirred only on the largest scales \footnote{\noindent
 A different stirring
  force is discussed in \cite{gnlo92a}.} and, of course, for these
large scales we have neglected viscosity and diffusivity.
$\u (\k) $ and $\Theta (\k )$ are the discrete Fourier
transforms of the velocity and the temperature fields (having
dimensions of length/time and temperature, respectively).

In anisotropic flow, a {\it correction} to (\ref{eq1}) is expected.
Lumley \cite{lum67} was the first to suggest the spectrum
\begineq
E_u (k) \sim \epsu^{2/3} k^{-5/3} \left(
1+ \alpha_u \left( {k\over k_s } \right)^{-2/3} \right)
\label{eq3}
\endeq
for an anisotropic flow, e.g., a shear flow. Here,
$\alpha_u$ is
an unknown dimensionless parameter
and $k_s$
the crossover wave vector separating the shear dominated range (SDR) and
the classical scaling range.
Eq.\ (\ref{eq3}) has also been obtained
by dimensional analysis in terms of Clebsch variables
\cite{kuz81,gro94,yak94}.

In what follows we will be interested in situations where the second term
in (\ref{eq3}) {\it dominates}. This definitely
is the case in cross spectra $E_{12}(k)$,
which probe anisotropy. Indeed, $E_{12}(k) \propto k^{-7/3}$ was
measured in boundary layer flow \cite{sad94}. Also,
Maloy and
Goldburg's \cite{mal93} scaling of the velocity structure function
$D^{(2)}(r) \propto r^{4/3} $ in Taylor--Couette flow with an
oscillatory
inner cylinder might
possibly be understood in terms of (\ref{eq3}) with a large $\alpha_u$.

We also mention that $E_u(k)\propto k^{-7/3}$ scaling is expected in
{\it helical} flow, i.e., flow for which the helicity
$\int \u \cdot \nab \times \u d\x $ is non zero.
The corresponding flux is then of order $T_{Hel}(k) \sim k^2 u^3(k)$.
Assuming that $T_{Hel}(k) = \epshel$ does not depend on $k$ in a
helicity dominated range, we obtain
\begineq
E_u(k)  \sim \epshel^{2/3}  k^{-7/3}.
\label{eq2}
\endeq
The crossover $k_s$ to classical scaling (\ref{eq1}) can be estimated
as $k_s=\epshel/\epsu$.

\section{Temperature shear spectra}
How does the spectrum of a passive scalar (e.g., the temperature) look,
if advected by a shear velocity field? From the conserved flux
$T_\Theta(k)\sim k u(k) \Theta^2(k) \sim \epst $ and the shear
velocity scaling $u(k) \sim \epsu^{1/3}k_s^{1/3}k^{-2/3}$ we readily conclude
$\Theta(k) \sim \epst^{1/2}\epsu^{-1/6}k_s^{-1/6}k^{-1/6}$ or
\begineq
E_\Theta(k)  \sim\epst \epsu^{-1/3}k_s^{-1/3}  k^{-4/3}.
\label{eq4}
\endeq
Considering also the high-$k$ range of the spectrum as in
(\ref{eq3}), we may write
\footnote{\noindent Strictly speaking, instead of the
    factor in brackets, one should write $(1+\alpha_u
    (k/k_s)^{-2/3})^{-1/2}$, and $\alpha_u$ and $\alpha_\Theta$ will
    not be independent.}
\begineq
E_\Theta (k) \sim \epst \epsu^{-1/3} k^{-5/3} \left(
1+ \alpha_\Theta \left( {k\over k_s } \right)^{-1/3} \right)^{-1}.
\label{eq5}
\endeq
The dimensionless constant $\alpha_\Theta$ will be estimated below.

Experimental support for our suggested scaling law (\ref{eq4})
is given by Sreenivasan \cite{sre91}, who
measured the power spectrum $\pto$ for temperature fluctuations in the
wake of a heated cylinder for $\rel = 200$. The power spectrum $\pto$
corresponds to $E_\Theta (k)$ via Taylor's hypothesis \cite{tay38}.
Indeed, $\pto \propto \om^{-4/3}$ for more than 1.5 decades, see fig.\
1. The long scaling range is surprising for the small $\rel$ as
it was in \cite{mal93}.
Unfortunately, the velocity spectrum in this experiment \cite{sre91}
was not measured. For the cross spectrum one would expect
$E_{12}(k)\propto k^{-7/3}$ as in the boundary layer \cite{sad94}.
We thus felt justified to assume $u(k)\propto k^{-2/3}$ here, as the flow
behind the cylinder is strongly anisotropic.

Where should we expect the crossover
between the SDR and the K41 range?
We estimate the crossover frequency
$\om_s$ directly by
the shear $s$. With the data given in \cite{sre91}, we find
$\om_s=s=U/L=885 sec^{-1}$. Here, $U=1770 cm/s$ is the flow velocity
and $L=2cm$ the diameter of the heated cylinder. Our result
$\log_{10}(\om_ssec)=2.9$ is indeed very near the end of the $\pto
\propto \om^{-4/3}$ scaling range, cf.\ fig.\ 1.
 The K41 subrange in between the SDR
and the viscous subrange (VSR) seems to be small, if it exists at all
for this small $\rel$.
Yet for larger $\rel$ it might show up in between SDR and VSR. Along
this line of argument we can understand why the (effective)
scaling exponent of $\pto$ approaches the classical
value $-5/3$, when $\rel$ is increased, see fig.\ 6 of ref.\
\cite{sre91}.

To be more quantitative, we compare the $\rel$
dependence of $k_s$ and that of the viscous dissipation cutoff $k_d$.
The latter is calculated to be \cite{amg94b} $k_d^{-1}\approx 18 L
\rel^{-3/2}$. For the former we
estimate $k_s^{-1} {\aleq} L$,
as in this geometry
shear is induced in the flow at some large length scale.
So for increasing $\rel$, a classical scaling range can develop between
the SDR $k< \alpha_\Theta^3 k_s$ and the VSR $k>k_d$, which
dominates the spectrum at large $k$.
{}From the experimental observation that
the K41 subrange is still missing for $\rel=200$ \cite{sre91}, we can
estimate $\alpha_\Theta$ to be $\alpha_\Theta
=(k_d(\rel=200)/k_s)^{1/3}\approx 5$. The classical scaling range will
grow as $k_d/(\alpha_\Theta^3k_s)=(\rel/200)^{3/2}$. For $\rel=1000$
it is already larger than one decade and more or less classical
scaling exponents will be extracted from the temperature spectra,
which accounts for the measurements summarized in
fig.\ 6 of \cite{sre91}.

\section{Rayleigh-Benard flow}
It is tempting to apply our considerations also to Rayleigh-Benard
(RB) flow, which has recently been studied in detail
[13--19].
RB flow is governed by a large scale ``wind''
\cite{cas89,til93,bel93}, i.e., the temperature is subjected to a
strong shear.
This wind is of course caused by buoyancy and somehow a reminiscence
of the convection rolls for lower $Ra$.
Yet it may well
be that the temperature in the interior can be considered as a
{\it passive} scalar. This view is supported by the observation,
that the local heat flux in the middle of the cell is tiny
\cite{til93,bel93,ben94b}.

First, we compare our predictions for the scaling exponents in shear
flow with the experimentally measured ones and with those of the
Bolgiano-Obukhov (BO59) scaling theory \cite{bol59},
which has become the canonical way to account for the RB scaling
exponents in the last years \cite{pro89},
but which has also been criticized \cite{shr90,gnlo91,gnlo92a,sig94}.
As we see from table 1, the experimental scaling exponents, with their
limited accuracy, neither favor the shear flow nor the BO59
explanation. Also the numerical simulations by Benzi et al.\
\cite{ben94b}, though excluding a pure K41 scenario, may not
be able to distinguish
between the BO59 and the shear scenario.
The numerical  simulations by Kerr and Malevsky
\cite{ker94} (3D) and by Werne \cite{wer93}
(2D) do not favor BO59.

One can also speculate whether RB flow can be considered as {\it helical}
flow (for a {\it single} realization due to spontaneous symmetry breaking)
so that (\ref{eq2}) can directly be applied. The effective forcing
of the velocity field due to the heating will partly be {\it helical}
so that, besides the energy flux, a helicity flux will be induced at
large scales, resulting in a $\propto k^{-7/3}$ spectrum as
discussed in Section 1.

In the following we will analyze whether the crossover predictions of
the two theories allow us to distinguish between them. Therefore we
shortly repeat the main ideas of BO59, which was originally
suggested for {\it stably  stratified} flow \cite{bol59}.
The kernel of the BO59 argument is dimensional analysis.
Assuming that
$\beta g$ and  $\epst$ are the relevant parameters, where
$\beta$ and $g$ are the volume expansion coefficient and the
acceleration due to gravity,  one obtains
\cite{bol59,my75,pro89,gnlo91}
\begineq
E_u(k)  \sim \epst^{2/5} (\beta g)^{4/5} k^{-11/5}, \qquad
E_\Theta(k) \sim \epst^{4/5} (\beta g)^{-2/5} k^{-7/5}
\label{eq7}
\endeq
for the velocity and temperature  wave vector spectra. If, instead,
$\epst$ and the kinetic energy dissipation rate $\epsu$ are the
relevant parameters, one retrieves Kolmogorov's classical result
(\ref{eq1}).
The crossover from BO59 to K41 is expected for wave vectors $k>k_B$,
where the
Bolgiano scale $k_B$  \cite{bol59,my75} is
obtained by equating (\ref{eq1}) and (\ref{eq7}),
\begineq
k_B = \epsu^{-5/4} \epst^{3/4} (\beta g)^{3/2}.
\label{eq8}
\endeq

With the exact result \cite{shr90,sig94}
\begineq
\epsu = \kappa^3 L^{-4} Pr Ra (Nu-1), \qquad
\epst = \kappa \Delta^2 L^{-2} Nu,
\label{eq9}
\endeq
the Bolgiano scale $k_B$ can be calculated. Here,
$Nu =
H/ (\kappa \Delta L^{-1})$ is the
Nusselt number, $H$ the heat flux
($=\langle u_3\Theta\rangle $ in the non diffusive case),
$Pr=\nu /\kappa$ the Prandtl number,
$Ra=\beta g \Delta L^3/(\nu\kappa )$ the Rayleigh number,
$L$
the height of the Rayleigh-Benard cell, $\Delta$
the temperature difference between the top and the bottom plate,
$\nu$ the viscosity, and $\kappa$ the diffusivity.
For large $Nu$
we obtain
from eq.\ (\ref{eq8}) $k_B^{-1} = Pr^{-1/4} Ra^{-1/4} Nu^{1/2} L$. The K41
range in the spectra should show up in between the BO59 range and the
viscous range, if $k_B^{-1}$ is larger than the typical crossover
scale to the viscous subrange. For the velocity structure function
 this scale is
known \cite{my75,gnlo93a} to be about $10\eta$, where $\eta=
\nu^{3/4} \epsu^{-1/4} = Pr^{1/2} Ra^{-1/4} Nu^{-1/4}L$
is the Kolmogorov length \cite{my75}.
For temperature structure functions and {\it high} $Pr$
 one could argue that $k_B^{-1}$ should be
compared with $10\eta_\Theta$, where $\eta_\Theta = Pr^{-1/2}\eta$ is the
thermal Kolmogorov length \cite{my75,bat59}, which
characterizes the crossover between the Batchelor regime (predicted
for  high $Pr$ in a passive scalar intensity spectrum
\cite{my75,bat59}) and the VSR. For very {\it low} Prandtl number
$Pr^{-3/4} \eta$ should be used \cite{my75,bat59}.
For existing experiments the resulting differences in the $Pr$
dependence cannot be detected anyhow as $Pr$ is always roughly 1, so
this difference is only of theoretical interest for the moment.
In what follows we
will compare $k_B^{-1}$ with $10\eta$ and give the $Pr$ dependences
resulting from a comparison of $k_B^{-1}$ with $10\eta_\Theta$
(i.e., the large $Pr$ case) in
brackets.

The condition $k_B^{-1} > 10\eta$ for the appearance of the K41 range
in between the BO59 and the viscous range requires that
\begineq
Nu> 10^{4/3} Pr \approx 22 Pr
\label{eq10}
\endeq
[$22 Pr^{1/3}$ when using $\eta_\Theta$ instead of $\eta$].
I.e., beyond a certain crossover Rayleigh number $Ra_{CO}$, defined by
$k_B^{-1} = 10\eta$, there will be {\it three } ranges.
For the Helium ($Pr=0.7$) convection experiment \cite{pro91}, $Nu=0.165
Ra^{2/7}$, or, embodying the
suggested {\footnote{\noindent The Shraiman-Siggia theory \cite{shr90,sig94}
assumes $Pr \gg 1$, but it has proven to work also for smaller $Pr$. Yet
all predictions on $Pr$ dependences should be taken with care.}}
 \cite{cas89,sig94} $Pr$
dependence, $Nu=0.157 Pr^{-1/7} Ra^{2/7}$.
Thus, if the BO59 scenario were true, pure BO59
scaling should be seen only for
$Ra < Ra_{CO} \approx 3\cdot 10^7
Pr^4\approx 10^7$ [$3\cdot 10^7 Pr^{5/2}$ for $\eta_\Theta $ instead of
$\eta$],
whereas for
$Ra > Ra_{CO} \approx  10^7$ also the K41 range should be visible.
Instead, pure $\ptos$ scaling is observed at least
up to $Ra\approx 10^{11}$
\cite{pro91}.
\footnote{\noindent
The nonuniversal spectral behavior for large $\omega$ and
$Ra> 10^{11}$ \cite{pro91} could consistently be accounted for
by considering the
reduced high frequency resolution of  the temperature probe
because of its finite size \cite{gnlo93a}.}
The discrepancy between $Ra_{CO}$ and $10^{11}$
can only  be resolved  by allowing for some
prefactors in the dimensional relations like eq.\ (\ref{eq8}) which are far
beyond the order of 1.
Note that $Ra_{CO}$ is near the experimentally observed
transition between soft and hard turbulence \cite{hes87,cas89,pro91}.
Our results predict that
only for larger $Pr$ the BO59 scaling range might become more extended.
Note that in this kind of analysis
we neglected the height dependence of $\eps$ and $\epst$ and thus
of $k_B$ by employing eqs.\ (\ref{eq9})\footnote{\noindent
In the numerical
simulations of Benzi et al.\ \cite{ben94b}, a height dependence
of $k_B$ is revealed. These simulations employ periodic boundary
conditions. BO59 scaling is found near the boundary (where $k_B$ is
large), but in contrast to experiment \cite{pro91}
not in the center of the cell.}.

For comparison, we now try to estimate the crossover between SDR and
K41 in RB flow. This amounts to a comparison between the typical shear scale
\begineq
k_s^{-1}= \epsu^{1/2}/s^{3/2}
\label{eq11}
\endeq
and $10\eta$.

First, we estimate the shear $s$ in RB flow. We do so
by using the
experimental information on the large scale velocity
$U=0.16 \kappa L^{-1} Ra^{1/2}$
\cite{pro91}
and on the mixing layer length $\lambda_m$, namely,
$s =U/\lambda_m$. Originally the mixing layer has been
theoretically introduced \cite{cas89} and later visualized
\cite{zoc90} as a region between the thermal boundary layer and
the bulk of turbulence. $\lambda_m/L \propto
Ra^{-1/7} $ is
found \cite{cas89,pro91}.
Embodying the predicted
$Pr$ dependence \cite{sig94},
this can be extended to
$\lambda_m / L \propto  Pr^{4/7}Ra^{-1/7} $.
In recent experiments \cite{til93,bel93}
$\lambda_m$ was redefined as distance of the maximum of the large scale
velocity to the wall.
In the employed $L=15.2 cm$ RB cell
\cite{til93,bel93}, a reduced high frequency resolution of
the temperature probe
(which is employed to measure $\lambda_m$, for the tricky
experimental details, see ref.\ \cite{bel93})
should be expected around $Ra_a =
(100 L^2/a^2)^{4/3}$ \cite{gnlo93a}, where $a=200\mu m$ is the
extension of the probe. Indeed, for $Ra < Ra_a$ the
measured \cite{bel93} mixing length
$\lambda_m$ agrees \cite{leo93} with the suggested $Ra$
scaling law $\lambda_m
\approx 1.2 L Pr^{4/7} Ra^{-1/7}$. We can thus finally estimate
\begineq
s = U/\lambda_m \approx  0.1 \kappa L^{-2} Pr^{-4/7} Ra^{9/14}.
\label{eq12}
\endeq

We now plug (\ref{eq12}) and (\ref{eq9}) into  (\ref{eq11})
and obtain
$k_s^{-1} \approx 8 L Pr^{9/7} Ra^{-9/28}$. When comparing $k_s^{-1}$ to
the crossover
$10\eta$
[or $10\eta_\Theta$] to the dissipation dominated range,
we get the remarkable result, that
the $Ra$ scaling of $k_s^{-1} $ and $10\eta$ is the same,
namely
\begineq
k_s^{-1} / 10\eta \approx  0.5 Pr^{3/4}
\label{eq13}
\endeq
[$0.5 Pr^{5/4}$ for $\eta_\Theta$ instead of $\eta$]
for {\it all} $Ra$. I.e., for such small $Pr$ as in the He cell ($Pr=0.7$)
the shear dominated range
is directly followed by the dissipation range with no K41 range in
between. For larger $Pr$ the K41 range might be observable.

Although the shear theory crossover scenario seems to agree better
with experiment than the one for BO59 (for small $Pr =0.7$),
many questions remain open and
our analysis is far from conclusive, as it does not go beyond
dimensional
analysis. Moreover, the estimation
$s=U/\lambda_m$ in eq.\ (\ref{eq12}) can be criticized, because
the shear might be
less in the center of the cell where $\pto$ is measured.
Also, the $z$-dependences of the energy dissipation rate
$\epsu$ and of the shear $s$ have  been
neglected.
The reason for this section is only  to demonstrate
that the BO59 scenario is not completely conclusive, either and that
the shear scenario might well be able to account for the experimental
observations.

\section{Summary}
To summarize, we derive the low frequency $\pto\propto
\om^{-4/3}$ power law for passive scalars
in a shear flow \cite{sre91}. It should be followed by a K41 range (for
high degree of turbulence) and of course by a viscous range. This
scenario might also account for the measured temperature power spectra
in RB flow \cite{pro91}, possibly without the K41 range in between SDR
and VSR for low $Pr$.
One of many open questions is the manner in which
higher-order moments scale. Benzi
et al.'s \cite{ben94b} numerics show strong intermittent behavior, so
dimensional analysis would fail here.
We hope that this work motivates the analysis of shear effects for both the
velocity and the temperature field in numerical simulations and in
experiment.

\vspace{1.5cm}
\noindent
{\bf Acknowledgements:}
We thank R.\ Almgren, R.\ Benzi,
F.\ Cattaneo,  G.\ Falkovich,
S.\ Grossmann,
L.\ Kadanoff, A.\ M\"uller-Groeling,
B.\ Shraiman, J.\ Wang, and J.\ Werne for discussions
and A.\ Tilgner and V.\ Yakhot for supplying us with the results of
their work prior to publication. The author also thanks the Aspen
Center of Physics for its hospitality.
Support by a NATO grant through
the Deutsche Akademische Austauschdienst (DAAD),
and by DOE is acknowledged.

\vspace{3cm}
\centerline{\bf Table}

 \begin{table}[htp]
 \begin{center}
 \begin{tabular}{|r|r|r|r|r|}
 \hline
         $   $
       & K41
       & BO59
       & shear flow
       & experiment
 \\
 \hline
         $P_u(\om)$
       & $-5/3$
       & $-11/5=-2.2$
       & $-7/3=-2.33$
       & $-2.2\pm 0.2$ \cite{ton92}
 \\
         $\pto$
       & $-5/3 $
       & $-7/5=-1.4$
       & $-4/3=-1.33$
       & $-1.3\pm 0.1$ \cite{pro91}
 \\
 \hline
 \end{tabular}
 \end{center}
 \end{table}

\vspace{1cm}

\noindent
Comparison of the scaling exponents predicted by different theories.
The last column shows the experimental results. Tong and Shen's
\cite{ton92} refer to velocity structure functions at $Ra=10^9 - 10^{10}$. I
estimated the experimental error to be at least $0.2$. Note that the
very precisely measured $\pto$ scaling exponent \cite{pro91} better
agrees with the shear flow scenario than with BO59.

\vspace{2cm}

\centerline{\bf Figure}


\begin{figure}[htb]
\caption[]{
Temperature power spectra in shear dominated flow,
measured behind
a wake of a heated cylinder. The data are
taken from Sreenivasan's work \cite{sre91} with kind permission of
the author.
The  spectrum shows a $-4/3$ power law. The dashed line shows the shear
crossover frequency $s$. In the inset the power spectrum, multiplied
by $\omega^{4/3}$, is  shown.
}
\label{fig1}
\end{figure}

\newpage


\noindent
$*$ On leave of absence from Fachbereich Physik, Universit\"at
Marburg,
Renthof 6, D-35032 Marburg.

\end{document}